\documentclass[5p,times]{elsarticle}
\usepackage{mathrsfs,bm}
\usepackage{longtable,lscape}
\usepackage{txfonts}
\usepackage{indentfirst}
\usepackage{graphicx,color,dcolumn,booktabs}

\usepackage{amsmath}

\usepackage{color}

\definecolor{cover}{rgb}{0.77,0.87,0.88}
\definecolor{blueone}{rgb}{0.1,0.1,.7}
\definecolor{citec}{rgb}{0.14,0.47,0.09}
\definecolor{two}{rgb}{0.0,0.5,0.}
\definecolor{three}{rgb}{.5,.1,0.15}
\usepackage[bookmarks=true,bookmarksopen=false,plainpages=false,breaklinks=true,
   bookmarksnumbered=true,hypertexnames=false,
   filecolor=blue,urlcolor=three,menucolor=three,
   linkcolor=three,citecolor=blueone, colorlinks,
   anchorcolor=blue,runcolor=pink,frenchlinks=red
   pdfstartview=FitH,pdftitle=title,%
   pdfauthor=author]{hyperref}

\allowdisplaybreaks[4]

\begin{document}

\begin{frontmatter}
\title{Possible $\Lambda_c\bar{\Lambda}_c$ molecular states and their productions in nucleon-antinucleon collision}
\author{Lin-Qing Song}
\author{Dan Song}
\author{Jun-Tao Zhu}
\author{Jun He}\ead{Corresponding author: junhe@njnu.edu.cn}
\address{School of Physics and Technology, Nanjing Normal University, Nanjing 210097, China}

\begin{abstract}

In this work, a study of possible molecular states from the
$\Lambda_c\bar{\Lambda}_c$ interaction and their productions in
nucleon-antinucleon collision is performed in a quasipotential Bethe-Salpeter
equation approach. Two bound states with quantum numbers $J^{PC}=0^{-+}$ and
$1^{--}$ are produced  with almost the same binding energy  from the
$\Lambda_c\bar{\Lambda}_c$ interaction which is described by the light meson
exchanges.  However, the result does not support the assignment of 
experimentally observed $Y(4630)$ as a $\Lambda_c\bar{\Lambda}_c$ molecular
state because it is hard to obtain a peak  near  experimental mass of the
$Y(4630)$ which is  far above the $\Lambda_c\bar{\Lambda}_c$ threshold.  The
possibility to search these states in nucleon-antinucleon collision is studied
by including couplings to $N\bar{N}$ and $D^{(*)}\bar{D}^{(*)}$ channels.   The
peaks can be found obviously near the $\Lambda_c\bar{\Lambda}_c$  threshold in
the $D^*\bar{D}^*$ channel at an order of amplitude of 10~$\mu$b.  Too small
width of state with $0^{-+}$ may lead to the difficulty to be observed in
experiment.  Based on the results in the current work, search for the
$\Lambda_c\bar{\Lambda}_c$ molecular state with $1^{--}$  is suggested in
process $N\bar{N}\to D^*\bar{D}^*$, which is accessible at $\rm \bar{P}ANDA$.

\end{abstract}

\begin{keyword}
  Molecular state \sep $\rm \bar{P}ANDA$ \sep $\Lambda_c\bar{\Lambda}_c$ interaction \sep Nucleon-antinucleon collision \sep Quasipotential Bethe-Saltpeter equation 
  \end{keyword}

  \end{frontmatter}

\section{Introduction}{\label{Introduction}}

The $Y(4630)$ with quantum numbers $J^{PC}$=$1^{--}$ was firstly observed in the
exclusive process $e^+e^-$$\to$$\Lambda_c \bar \Lambda_c$ by Belle
collaboration~\cite{Belle:2008xmh}, and confirmed at BESIII after several
years~\cite{BESIII:2017kqg}. It carries mass and width of
$M$=4634$_{-7}^{+8}$(stat.)$^{+5}_{-8}$(sys.)~MeV and
$\Gamma_{tot}$=92$_{-24}^{+40}$(stat.)$^{+10}_{-21}$(sys.)~MeV, respectively.
The internal structure of $Y(4630)$ attracts attentions from the community  and
many interpretations have been proposed to understand its internal structure. It
was explained as a conventional charmonium state in
Refs.~\cite{Badalian:2008dv,Segovia:2008ta,Xiao:2018iez,Wang:2020prx}. The
$Y(4630)$  as a tetraquark was also supported by calculation in the constituent
quark model and QCD sum rule~\cite{Liu:2016sip,Wang:2018rfw}.  Since the mass of
$Y(4630)$ is close to the $\Lambda_c \bar{\Lambda}_c$ threshold, this structure
was also related to the threshold effect~\cite{vanBeveren:2008rt} and $\Lambda_c
\bar{\Lambda}_c$ baryonium~\cite{Chen:2011cta}. Meanwhile, the
$\Lambda_c\bar{\Lambda}_c$  system also attracts
attentions~\cite{Wang:2021qmn,Lee:2011rka,Simonov:2011jc}. Theoretical
calculations suggest strong attraction between a $\Lambda_c$ baryon and a
$\bar{\Lambda}_c$ baryon by $\sigma$ and $\omega$ exchanges, which favors the
existence of a $\Lambda_c \bar{\Lambda}_c$ molecular
states~\cite{Lee:2011rka,Simonov:2011jc}.

It is worthwhile noting that $Y(4630)$ is near another exotic state $Y(4660)$,
which was observed in the invariant mass spectrum of
$\psi(2S)\pi^+\pi^-$~\cite{Belle:2007umv}. Within experimental uncertainties,
the two states were reckoned as the same
state~\cite{Bugg:2008sk,Cotugno:2009ys}.  It was also proposed that the
$Y(4660)$ is more likely to be a molecular state of
$\psi(2S)f_0(980)$~\cite{Guo:2008zg}.  In Ref.~\cite{Guo:2010tk},  the
possibility that the Y(4630) peak is of the same origin as the $Y(4660)$ was
discussed.  Ref.~\cite{Dong:2021juy} questions the existence of the $Y(4630)$,
and discusses that the vector $\Lambda_c \bar{\Lambda}_c$ bound state predicted
could have its signal in the BESIII measurement~\cite{BESIII:2017kqg}. 

If we put the two states together, it seems that the mass of $Y(4630)$ favors
the  assignment of $Y(4630)$ as a $\Lambda_c \bar{\Lambda}_c$ molecular state.
But there is an obvious difficulty to fit the experimental peak of $Y(4630)$
with a $\Lambda_c \bar{\Lambda}_c$ molecular state.  The $Y(4630)$ is about 60
MeV above the $\Lambda_c \bar{\Lambda}_c$ threshold  while a molecular state is
usually assumed as a bound state with mass below the threshold.  To solve this
problem, it was suggested that the data from $\Lambda_c \bar{\Lambda}_c$
threshold up to 4.7 GeV should contain signals of at least two states,  a
$\Lambda_c \bar{\Lambda}_c$ molecule and another one with a mass around 4.65
GeV~\cite{Dong:2021juy}. 

To make clear the internal structure of the $Y(4630)$ and the structures near
the $\Lambda_c\bar{\Lambda}_c$ threshold, more theoretical and experimental
studies are required. Since a $\Lambda_c\bar{\Lambda}_c$ molecular state is a
baryonium, it should  be easy to be produced from a pair of a nucleon and an
antinucleon at high energy collision by exciting  a charm-anticharm quark pair.
Such process is accessible in the $\rm \bar{P}ANDA$ (AntiProton annihilations at
Darmstadt) detector at FAIR (Facility for Antiproton and Ion Research) Recently,
some theoretical researches about $\Lambda_c \bar{\Lambda}_c$  production
through proton-antiproton collision were performed  in the literature, and
considerably large cross section was
predicted~\cite{Titov:2008yf,Goritschnig:2009sq,Haidenbauer:2009ad,Wang:2016fhj,Guo:2016iej,Wiedner:2011mf,Khodjamirian:2011sp}.
The charm-meson pair production from the proton-antiproton collision were also
studied in the literature~\cite{Titov:2008yf,Shyam:2016bzq,Shi:2021hzm}.
However, in these studies, the direct relation between the molecular state and
the production process was not considered.

In the current work, the molecular states from the $\Lambda_c\bar{\Lambda}_c$
interaction will be studied in a quasipotential Bethe-Salpeter equation (qBSE)
approach~\cite{He:2019ify,Zhu:2019ibc,He:2019csk}. The
$\Lambda_c\bar{\Lambda}_c$ invariant mass spectrum will be estimated to discuss
the possibility of assignment of the $Y(4630)$ as a $\Lambda_c \bar{\Lambda}_c$
molecular state. As discussed above, a virtual or resonance state is required to
interpret the mass gap between the threshold and the experimental mass. It will
be seen that such requirement is difficult to be satisfied. If the $\Lambda_c
\bar{\Lambda}_c$ molecular state is not the $Y(4630)$, it is interesting to
study the possibility to search the obtained molecular states in the
nucleon-antinucleon collision.  In this work, the coupling of the molecular
states to the $N\bar{N}$ and $D^{(*)}\bar{D}^{(*)}$ channels will be constructed
and the cross sections of processes $N\bar{N} \to N\bar{N}$ and
$D^{(*)}\bar{D}^{(*)}$ will be calculated.

This article is organized as follows. After introduction, we present the details
of theoretical frame in Section~\ref{THEORETICAL FRAME}, which includes relevant
effective Lagrangians and parameters, the potential kernel of the $\Lambda_c
\bar{\Lambda}_c$ interaction  and coupled-channel interaction with $N\bar{N}$
and $D^{(*)}\bar{D}^{(*)}$ channels, and a brief introduction about the qBSE
approach.  The numerical results of the molecular states from the $\Lambda_c
\bar{\Lambda}_c$ interaction will be given in Section~\ref{NUMERICAL RESULTS}.
The relation between such states and Y(4630) will be discussed. The
coupled-channel calculation is also performed and the cross sections of the
processes $N\bar{N} \to N\bar{N}, D^{(*)}\bar{D}^{(*)}$ are also presented in
this section. Finally, article ends with a summary in section~\ref{Summary}.

\section{Theoretical frame}\label{THEORETICAL FRAME}

\subsection{Interaction mechanism and relevant Lagrangians}

\begin{figure}[h!]
  \includegraphics[bb=90 360 470 762,clip,scale=0.6]{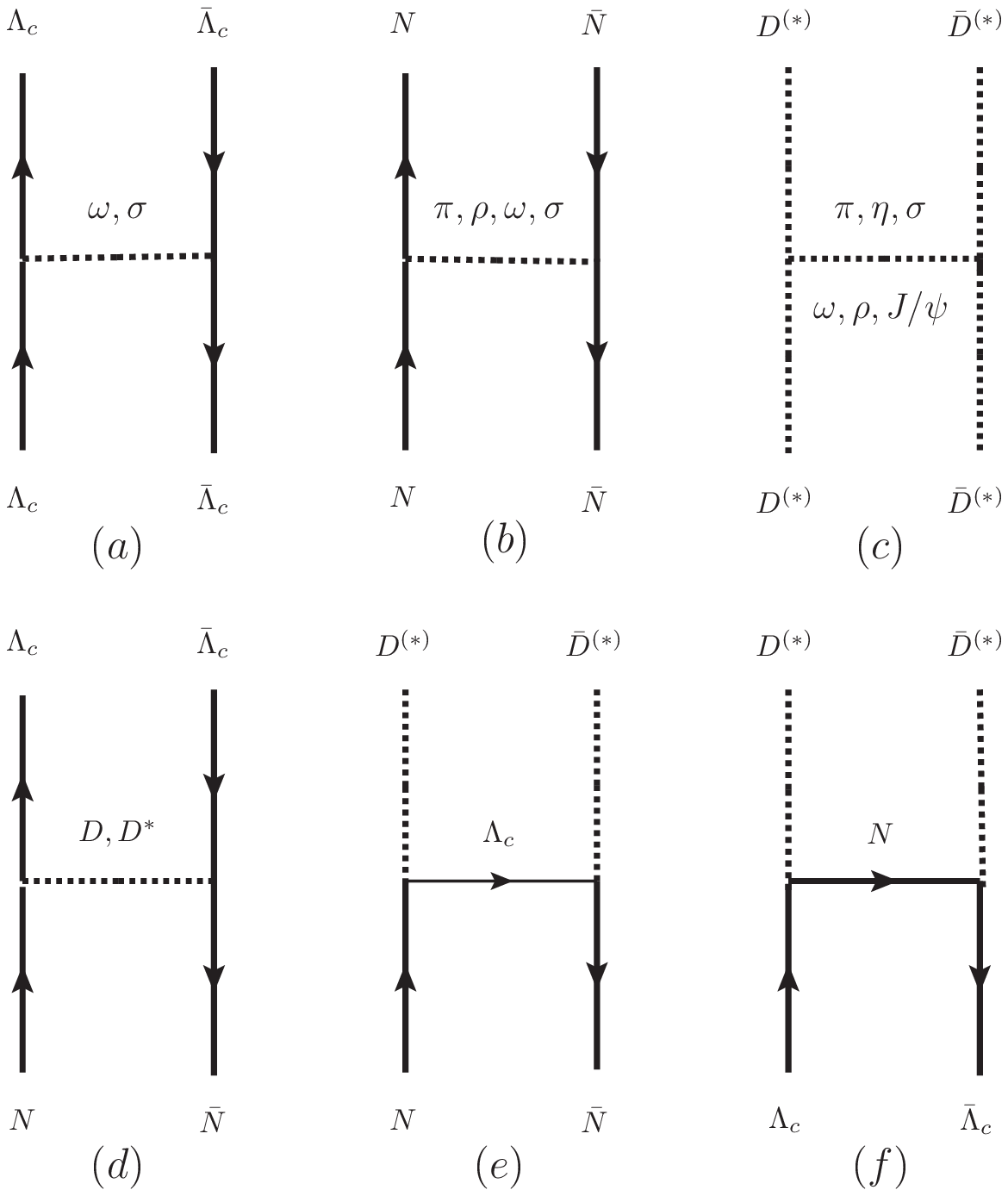}
  \caption{Diagrams for the interactions involved in the current work including single channels $\Lambda_c\bar{\Lambda}_c$ (a), $N\bar{N}$ (b),  and  $D^{(*)}\bar{D}^{(*)}$ (c), and coupled channels $N\bar{N}\to\Lambda_c\bar{\Lambda}_c$ (d), $N\bar{N}\to D^{(*)}\bar{D}^{(*)}$ (e), and $\Lambda_c\bar{\Lambda}_c\to D^{(*)}\bar{D}^{(*)}$ (f). }
  \label{feyn}
  \end{figure}

In the current work, we focus on the molecular states from the $\Lambda_c
\bar{\Lambda}_c$ interaction, which mechanism is described in Fig~\ref{feyn}
(a). Since the $\Lambda_c$ baryon is isoscalar state, only the isoscalar $\omega$ and
$\sigma$ mesons can be exchanged to provide the interaction.

To write the potential for the $\Lambda_c \bar{\Lambda}_c$ interaction, the
Lagrangians constructed with heavy quark and chiral symmetries are introduced, and presented  explicitly as follows~\cite{Yan:1992gz,Cheng:1992xi,Casalbuoni:1996pg, He:2019rva,Zhu:2021lhd},
\begin{align}
{\cal L}_{\Lambda_c\Lambda_c\mathbb{V}}=-i\frac{g_V\beta_B}{4m_{\Lambda_c}}\omega^\mu\bar{\Lambda}_c\overleftrightarrow{\partial}_\mu\Lambda_c,\quad
{\cal L}_{\Lambda_c\Lambda_c\sigma}=i\ell_B \sigma \bar{\Lambda}_c \Lambda_c,
   \label{LB}
\end{align}
where the $\Lambda_c$, $\omega$ meson, and $\sigma$ are for $\Lambda_c$ baryon, $\omega$, and $\sigma$ meson fields, $m_{\Lambda_c}$ is the mass of $\Lambda_c$ baryon. The coupling constants are chosen as $g_V=5.9$, $\beta_B=0.87$, $\ell_B=-3.1$~\cite{He:2019rva,Zhu:2021lhd}.

To study the production of the $\Lambda_c \bar{\Lambda}_c$  molecular states in
nucleon-antinucleon collision, more channels need to be introduced, that is, the
$N\bar{N}$ and $D^{(*)}\bar{D}^{(*)}$ interactions,  which are depicted with
Feynman diagrams in Fig~\ref{feyn} (b) and  Fig~\ref{feyn} (c).  The Lagrangians
for the vertices of coupling of the nucleon to  pseudoscalar meson $\pi$, vector
meson $\rho/\omega$, and $\sigma$ meson are~\cite{Ronchen:2012eg} \begin{align}
\mathcal{L}_{NN\pi}
&=-\frac{g_{NN\pi}}{m_{\pi}}\bar{N}\gamma^5\gamma^{\mu}\mathop{\tau}\limits^{\rightarrow}
\cdot\partial_{\mu}\mathop{\pi}\limits^{\rightarrow}N,\quad
\mathcal{L}_{NN\sigma} =-g_{NN\sigma}\bar{N}N\sigma,\nonumber\\
\mathcal{L}_{NN\rho} &=
-g_{NN\rho}\bar{N}[\gamma^{\mu}-\frac{\kappa_{\rho}}{2m_{N}}\sigma^{\mu\nu}\partial_{\nu}]
\mathop{\tau}\limits^{\rightarrow}\cdot\mathop{\rho_{\mu}}\limits^{\rightarrow}N,
\nonumber\\
\mathcal{L}_{NN\omega}
&=-g_{NN\omega}\bar{N}[\gamma^{\mu}-\frac{\kappa_{\omega}}{2m_{N}}\sigma^{\mu\nu}\partial_{\nu}]
\omega_{\mu}N,\label{LD} \end{align} where $N$, $\pi$, $\rho$, $\omega$ are
nucleon, pion meson, $\rho$ meson, and $\omega$ meson fields. The coupling
constants $g_{NN\pi}=0.989$, $g_{NN\rho}=-3.1$, $\kappa_{\rho}=1.825$,
$\kappa_{\omega}=0$, and $g_{NN\sigma}=5$~\cite{Ronchen:2012eg,Matsuyama:2006rp,Lu:2020qme}. The
coupling constants for the $\omega$ meson can be related to these for the $\rho$
meson with  SU(3) symmetry as $g_{NN\omega}=3g_{NN\rho}$.  In the current work,
baryon-antibaryon system is considered. The interaction of a baryon and an
antibaryon can be easily related to the interaction with two baryons with the
$G$-parity rule as $V_{B\bar{B}M}=(-1)^GV_{BBM}$ where G is the  G-parity of
exchanged meson $M$~\cite{Zhu:2019ibc,Phillips:1967zza}. So, we can relate the
nucleon-antinucleon interaction in the current work to the nucleon-nucleon
interaction, which can produce the deuteron. With such interaction,  the
experimental mass of deuteron can be reproduced in the current model. The
$G$-parity rule can be also applied to $\Lambda_c\bar{\Lambda}_c$ interaction.

The couplings of  heavy-light charmed mesons
$D^{(*)}=(D^{(*)0},D^{(*)+},D_s^{(*)+})$ and its antiparticle
$\bar{D}^{(*)}=(\bar{D}^{(*)0},D^{(*)-},D_s^{(*)-})$ to the light mesons can be depicted by the Lagrangians with heavy quark and chiral symmetries as~\cite{Yan:1992gz,Cheng:1992xi,Casalbuoni:1996pg},
\begin{align}\label{eq:lag-p-exch}
    \mathcal{L}_{D^*D\mathbb{P}} &=
    -i\frac{2g\sqrt{m_Dm_{D^*}}}{f_\pi} (-D_bD^{*\dag}_{a\lambda}+D^*_{b\lambda}D^\dag_{a})\partial^\lambda{}\mathbb{P}_{ba}
   \nonumber\\
   & +i\frac{2g\sqrt{m_Dm_{D^*}}}{f_\pi}
    (-\tilde{D}^{*\dag}_{a\lambda}\tilde{D}_b+\tilde{D}^\dag_{a}\tilde{D}^*_{b\lambda})\partial^\lambda\mathbb{P}_{ab},\nonumber\\
      \mathcal{L}_{D^*D^*\mathbb{P}} &=
    \frac{g}{f_\pi}\epsilon_{\alpha\mu\nu\lambda}D^{*\mu}_b\overleftrightarrow{\partial}^\alpha D^{*\lambda\dag}_{a}\partial^\nu\mathbb{P}_{ab}
    -\frac{g}{f_\pi} \epsilon_{\alpha\mu\nu\lambda}\tilde{D}^{*\mu\dag}_a\overleftrightarrow{\partial}^\alpha \tilde{D}^{*\lambda}_{b}\partial^\nu\mathbb{P}_{ba},\nonumber\\
    \mathcal{L}_{D^*D\mathbb{V}} &=
    \sqrt{2}\lambda g_V\epsilon_{\lambda\alpha\beta\mu}
    (-D^{*\mu\dag}_a\overleftrightarrow{\partial}^\lambda D_b
    +D^\dag_a\overleftrightarrow{\partial}^\lambda D_b^{*\mu})
    (\partial^\alpha{}\mathbb{V}^\beta)_{ba}\nonumber\\
    &+\sqrt{2}\lambda g_V\epsilon_{\lambda\alpha\beta\mu}
    (-\tilde{D}^{*\mu\dag}_a\overleftrightarrow{\partial}^\lambda
    \tilde{D}_b  +\tilde{D}^\dag_a\overleftrightarrow{\partial}^\lambda
    \tilde{D}_b^{*\mu})(\partial^\alpha{}\mathbb{V}^\beta)_{ab},\nonumber\\
    \mathcal{L}_{DD\mathbb{V}} &=
   -i\frac{\beta g_V}{\sqrt{2}} D_a^\dag\overleftrightarrow{\partial}^\mu D_b\mathbb{V}^\mu_{ba}+i\frac{\beta	g_V}{\sqrt{2}}\tilde{D}_a^\dag
  \overleftrightarrow{\partial}^\mu \tilde{D}_b\mathbb{V}^\mu_{ab},\nonumber\\
    \mathcal{L}_{D^*D^*\mathbb{V}} &=
   i\frac{\beta g_V}{\sqrt{2}}D_a^{*\dag}\overleftrightarrow{\partial}^\mu D^*_b\mathbb{V}^\mu_{ba}- i\frac{\beta g_V}{\sqrt{2}}\tilde{D}_a^{*\dag}\overleftrightarrow{\partial}^\mu\tilde{D}^*_b\mathbb{V}^\mu_{ab}
   \nonumber\\
   & -i2\sqrt{2}\lambda g_V m_{D^*}D^{*\mu}_bD^{*\nu\dag}_a(\partial_\mu\mathbb{V}_\nu-\partial_\nu\mathbb{V}_\mu)_{ba}
    \nonumber\\
    &- i2\sqrt{2}\lambda  g_Vm_{D^*}\tilde{D}^{*\mu\dag}_a\tilde{D}^{*\nu}_b(\partial_\mu\mathbb{V}_\nu-\partial_\nu\mathbb{V}_\mu)_{ab},\nonumber\\
    \mathcal{L}_{DD\sigma} &=
    -2g_\sigma m_{D}D_a^\dag D_a\sigma -2g_\sigma m_{D}\tilde{D}_a^\dag \tilde{D}_a\sigma,\nonumber\\
    \mathcal{L}_{D^*D^*\sigma} &=
    2g_\sigma m_{D^*}D_a^{*\dag} D^*_a\sigma +2g_\sigma m_{D^*}\tilde{D}_a^{*\dag}
    \tilde{D}^*_a\sigma.
  \end{align}
  Here we adopt the parameters as $f_{\pi}=132~$MeV, $g=0.59$, $\beta=0.9$, $\lambda$=0.56~GeV$^{-1}$, $g_V=5.9$, and $g_\sigma=0.76$~\cite{Yan:1992gz,Cheng:1992xi,Casalbuoni:1996pg}. The $\mathbb{V}$ and $\mathbb{P}$ are the vector and pseudoscalar matrices as
  \begin{align}
  {\mathbb P}=\left(\begin{array}{ccc}
          \frac{\sqrt{3}\pi^0+\eta}{\sqrt{6}}&\pi^+&K^+\\
          \pi^-&\frac{-\sqrt{3}\pi^0+\eta}{\sqrt{6}}&K^0\\
          K^-&\bar{K}^0&\frac{-2\eta}{\sqrt{6}}
  \end{array}\right),
  \mathbb{V}=\left(\begin{array}{ccc}
  \frac{\rho^{0}+\omega}{\sqrt{2}}&\rho^{+}&K^{*+}\\
  \rho^{-}&\frac{-\rho^{0}+\omega}{\sqrt{2}}&K^{*0}\\
  K^{*-}&\bar{K}^{*0}&\phi
  \end{array}\right).\nonumber
  \end{align}

  The  couplings between heavy-light mesons and $J/\psi$ are also required, which are of forms~\cite{Aceti:2014uea,He:2017lhy}
  \begin{align}
    {\cal L}_{D^*\bar{D}^*J/\psi}&=-ig_{D^*D^*\psi}
  \big[\psi \cdot \bar{D}^*\overleftrightarrow{\partial}\cdot D^*  \nonumber\\&-
  \psi^\mu \bar D^* \cdot\overleftrightarrow{\partial}^\mu {D}^* +
  \psi^\mu \bar{D}^*\cdot\overleftrightarrow{\partial} D^{*\mu} ) \big] ,\nonumber \\
  {\cal L}_{D^*\bar{D}J/\psi}&=
  -g_{D^*D\psi} \,  \, \epsilon_{\beta \mu \alpha \tau}
  \partial^\beta \psi^\mu (\bar{D}
  \overleftrightarrow{\partial}^\tau D^{* \alpha}+\bar{D}^{* \alpha}
  \overleftrightarrow{\partial}^\tau D) ,\label{matrix3} \nonumber \\
  {\cal L}_{D\bar{D}J/\psi} &=
  ig_{DD\psi} \psi \cdot
  {D}\overleftrightarrow{\partial}\bar{D},
  \end{align}
  where the coupling constants above satisfy the relation as ${g_{D^*D^*\psi}}/{m_{D^*}} = {g_{DD\psi}}/{m_D}=
  g_{D^*D\psi}= 2 g_2 \sqrt{m_\psi }$ and $g_2={\sqrt{m_\psi}}/{2m_Df_\psi}$ with $f_\psi=405$ MeV.

Finally, the three channels considered above can be coupled to each other as
shown in Fig~\ref{feyn}(d), Fig~\ref{feyn}(e), and Fig~\ref{feyn}(f) by the $D^{(*)}N\Lambda_c$ vertices,
the Lagrangians of which are constructed under the SU(4)
symmetry~\cite{Okubo:1975sc,
Wang:2016fhj,Sibirtsev:2000aw,Liu:2001ce,Dong:2009tg} as follows,
\begin{align}
{\cal L}_{DN\Lambda_c}&=ig_{DN\Lambda_c}(\bar{N}\gamma_5\Lambda_cD+\bar{D}\bar{\Lambda}_c\gamma_5N),\nonumber\\
{\cal L}_{D^*N\Lambda_c}&=g_{D^*N\Lambda_c}(\bar{N}\gamma_\mu\Lambda_cD^{*\mu}+\bar{D}^{*\mu}\bar{\Lambda}_c\gamma_{\mu}N),
\label{DNLc} \end{align}
where the coupling constants, $g_{DN\Lambda_c}$ and
$g_{D^*N\Lambda_c}$, are chosen as $10.7$ and $-5.8$,
respectively~\cite{Khodjamirian:2011sp}.

\subsection{Potential kernel and qBSE approach}\label{2.1}

In the current work, we focus on the $\Lambda_c\bar{\Lambda}_c$ molecular states
and their productions. Since baryons $\Lambda_c$ and $\bar{\Lambda}_c$ are
isoscalar states, only isoscalar flavor functions are considered
as~\cite{Dong:2021juy,Ding:2020dio},
\begin{align}
  |\Psi_{\Lambda_c\bar{\Lambda}_c}\rangle &=-|\Lambda_c^+ \Lambda_c^-\rangle, \quad
  |\Psi_{N\bar{N}}\rangle =\frac{1}{\sqrt{2}}\Big[|p\bar{p}\rangle+|n\bar{n}\rangle\Big],\nonumber\\
  |\Psi_{D\bar{D}}\rangle &=\frac{1}{\sqrt{2}}\Big[|D^+D^-\rangle+|D^0\bar{D}^0\rangle\Big],\nonumber\\
  |\Psi_{D\bar{D}^*}\rangle &=\frac{1}{2}\Big[\big(|D^{*+}D^-\rangle+|D^{*0}\bar{D}^0\rangle\big)+c\big(|D^+D^{*-}\rangle+|D^0\bar{D}^{*0}\rangle\big)\Big],\nonumber\\
  |\Psi_{D^*\bar{D}^*}\rangle &=\frac{1}{\sqrt{2}}\Big[|D^{*+}D^{*-}\rangle+|D^{*0}\bar{D}^{*0}\rangle\Big],
   \label{Eq: wf1}
\end{align}
where $c=\pm$ corresponds to $C$ parity $C=\mp$ respectively.

With the wave functions and Lagrangians, the potential can be obtained by
applying the standard Feynman rules in the one-boson-exchange model
as~\cite{He:2019ify},
\begin{align}%
  {\cal V}_{\mathbb{P},\sigma}&=I_i^{(d,c)}\Gamma_1\Gamma_2 P_{\mathbb{P},\sigma}f(q^2),\quad
  {\cal V}_{\mathbb{V}}=I_i^{(d,c)}\Gamma_{1\mu}\Gamma_{2\nu}  P^{\mu\nu}_{\mathbb{V}}f(q^2),\nonumber\\
  {\cal V}_{\mathbb{B}}&=I_i^{(d,c)}\Gamma_1P_{\mathbb{B}}\Gamma_2  f(q^2),
  \label{V}
\end{align}
where the $I_i^d$ ($I_i^c)$ is flavor factors for certain meson exchange of
certain interaction, which are calculated explicitly with the flavor functions and
Lagrangians, and the values are listed as Table~\ref{flavorfactor}.

\renewcommand\tabcolsep{0.24cm}
\renewcommand{\arraystretch}{1.2}
  \begin{table}[h!]\footnotesize 
  \begin{center}
  \caption{The flavor factors $I_i^d$ ($I_i^c)$ for direct and cross diagrams respectively and different exchange particles in corresponding channels.}
  \label{flavorfactor}
  \begin{tabular}{c|cccccc}\bottomrule[1pt]
     &$\pi$  &$\eta$         &$\rho$        &$\omega$      &$\sigma$ &$J/\psi$ \\\hline
  $N\bar{N}-N\bar{N}$    &$3$   &$--$&$-3$ &$-1$ & $1$    &$--$              \\
  $D\bar{D}-D\bar{D}$    &$--$   &$--$          &$\frac{3}{2}$ &$\frac{1}{2}$ & $1$     &$1$        \\
  $D\bar{D}^*-D\bar{D}^*$  &$(\frac{3}{2})$   &$(\frac{1}{6})$          &$\frac{3}{2}(\frac{3}{2})$ &$\frac{1}{2}(\frac{1}{2})$ &$1$      &$1(1)$  \\
  $D^*\bar{D}^*-D^*\bar{D}^*$&$\frac{3}{2}$&$\frac{1}{6}$&$\frac{3}{2}$&$\frac{1}{2}$&$1$   &$1$          \\
  $\Lambda_{c}\bar{\Lambda}_{c}-\Lambda_{c}\bar{\Lambda}_{c}$&$--$&$--$&$--$       &$-2$          & $4$     &$--$      \\
  $D\bar{D}-D\bar{D}^*$     &$--$ &$--$ &$\frac{3}{2}(\frac{3}{2})$ &$\frac{1}{2}(\frac{1}{2})$ & $--$   &$1(1)$   \\
  $D\bar{D}-D^*\bar{D}^*$&$--$&$--$&$\frac{3}{2}$&$\frac{1}{2}$&$--$   &$1$       \\
  $D \bar{D}^*-D^*\bar{D}^* $  &$\frac{3}{2}(\frac{3}{2})$ &$\frac{1}{6}(\frac{1}{6})$ &$\frac{3}{2}(\frac{3}{2})$ &$\frac{1}{2}(\frac{1}{2})$ & $--$   &$1(1)$  \\\hline
  &$D$        &$D^*$       &$N$       &$\Lambda_c$&\\\hline
  $N\bar{N}-D\bar{D}                      $         &$--$      &$--$&$--$      &$1$&\\
  $N\bar{N}-D\bar{D}^*                    $         &$--$     &$--$     &$--$      &$1(1)$&\\
  $N\bar{N}-D^*\bar{D}^*                  $       &$--$      &$--$     &$--$      &$1$&\\
  $N\bar{N}-\Lambda_{c}\bar{\Lambda}_{c}   $        &$-\sqrt{2}$      &$-\sqrt{2}$   &$--$      &$--$&\\
  $D\bar{D}-\Lambda_{c}\bar{\Lambda}_{c}   $        &$--$      &$--$  &$-\sqrt{2}$      &$--$&\\
  $D\bar{D}^*-\Lambda_{c}\bar{\Lambda}_{c}   $        &$--$      &$--$  &$-\sqrt{2}(-\sqrt{2})$      &$--$&\\
  $D^*\bar{D}^*-\Lambda_{c}\bar{\Lambda}_{c}$         &$--$      &$--$  &$-\sqrt{2}$      &$--$&\\
  \toprule[1pt]
  \end{tabular}
  \end{center}
\end{table}

The propagators for the exchanged pseudoscalar $\mathbb{P}$ meson, scalar $\sigma$  meson, vector $\mathbb{V}$ meson, and baryon $\mathbb{B}$ are defined as,
\begin{align}%
  P_{\mathbb{P},\sigma}= \frac{i}{q^2-m_{\mathbb{P},\sigma}^2},
  P^{\mu\nu}_\mathbb{V}=i\frac{-g^{\mu\nu}+q^\mu q^\nu/m^2_{\mathbb{V}}}{q^2-m_\mathbb{V}^2},
  P_{\mathbb{B}}=\frac{i(\not q+m)}{q^2-m^2}.\nonumber
\end{align}
We introduce the $f(q^2)$ as a form factor to compensate the off-shell effect of
exchanged meson, which can be written concretely as
$f(q^2)=e^{-(m_e^2-q^2)^2/\Lambda_e^2}$ with $m_e$ and $q$ being the mass
and the momentum of  exchanged meson. The cutoff is parameterized as 
\begin{align}
\Lambda_e=m_e+\alpha~0.22~{\rm GeV}.
\end{align}

After inserting the potential kernel into the Bethe-Salpeter equation to obtain
the scattering amplitude and applying spectator quasipotential approximation and
partial-wave  decomposition, a 1-dimensional integral equation with fixed
spin-parity $J^P$ can be obtained
as~\cite{He:2019ify,Zhu:2019ibc,He:2019csk,Ding:2020dio,Zhu:2021lhd}
\begin{align}
  i{\cal M}^{J^P}_{\lambda'\lambda}({\rm p}',{\rm p})
  &=i{\cal V}^{J^P}_{\lambda',\lambda}({\rm p}',{\rm
  p})+\sum_{\lambda''}\int\frac{{\rm
  p}''^2d{\rm p}''}{(2\pi)^3}\nonumber\\
  &\cdot
  i{\cal V}^{J^P}_{\lambda'\lambda''}({\rm p}',{\rm p}'')
  G_0({\rm p}'')i{\cal M}^{J^P}_{\lambda''\lambda}({\rm p}'',{\rm
  p}),\quad\quad \label{Eq: BS_PWA}
  \end{align}
where the ${\cal M}^{J^P}_{\lambda'\lambda}({\rm p}',{\rm p})$ is partial-wave
scattering amplitude and the $G_0({\rm p}'')$ is reduced propagator under
quasipotential approximation. 
The partial wave potential is defined with the potential
obtained  in Eq.~(\ref{V}) as
\begin{align}
  {\cal V}_{\lambda'\lambda}^{J^P}({\rm p}',{\rm p})
  &=2\pi\int d\cos\theta
  ~[d^{J}_{\lambda\lambda'}(\theta)
  {\cal V}_{\lambda'\lambda}({\bm p}',{\bm p})\nonumber\\
  &+\eta d^{J}_{-\lambda\lambda'}(\theta)
  {\cal V}_{\lambda'-\lambda}({\bm p}',{\bm p})],
\end{align}
where $\eta=PP_1P_2(-1)^{J-J_1-J_2}$ with $P$ and $J$ being parity and spin for
the $\Lambda_c\bar{\Lambda}_c$ system. The initial and final relative momenta
are chosen as ${\bm p}=(0,0,{\rm p})$  and ${\bm p}'=({\rm p}'\sin\theta,0,{\rm
p}'\cos\theta)$. The $d^J_{\lambda\lambda'}(\theta)$ is the Wigner d-matrix.  An
exponential regularization is introduced to reduced propagator $G_0({\rm
p}'')\to G_0({\rm p}'')\left[e^{-(p''^2_l-m_l^2)^2/\Lambda_r^4}\right]^2$ with
$\Lambda_r$ being the cutoff chosen as
$\Lambda_r=\Lambda_e$~\cite{He:2015mja}.

The molecular state from the $\Lambda_{c}\bar{\Lambda}_{c}$ interaction
corresponds to a pole of scattering amplitude $\cal M$. With discretizing the
momenta ${\rm p}$, ${\rm p}'$, and ${\rm p}''$ to ${\rm p}_i$, ${\rm p}'_i$, and
${\rm p}''_i$ with the weight factor  $w({\rm p}''_j)$ by the Gauss quadrature
the integral equation can be transformed to a  matrix equation as~\cite{He:2015mja},
\begin{eqnarray}
  {M}_{ik}
  &=&{V}_{ik}+\sum_{j=0}^N{ V}_{ij}G_j{M}_{jk}.
\end{eqnarray}
The discretized propagator is of a form
\begin{eqnarray}
    G_{j>0}&=&\frac{w({\rm p}''_j){\rm p}''^2_j}{(2\pi)^3}G_0({\rm
    p}''_j), \nonumber\\
  G_{j=0}&=&-\frac{i{\rm p}''_o}{32\pi^2 W}+\sum_j
  \left[\frac{w({\rm p}_j)}{(2\pi)^3}\frac{ {\rm p}''^2_o}
  {2W{({\rm p}''^2_j-{\rm p}''^2_o)}}\right],
\end{eqnarray}
with ${\rm p}''_o=\frac{1}{2W}\sqrt{[W^2-(M_1+M_2)^2][W^2-(M_1-M_2)^2]}$.   And
the pole can be searched by finding a complex energy $z=E_R+i\Gamma/2$  which
satisfies $|1-V(z)G(z)|=0$. Here $E_R$ and $\Gamma$ are mass
and decay width of the molecular state, respectively.  

The total cross section can be obtained as follows~\cite{He:2015cca},
\begin{eqnarray}
    \sigma=\frac{1}{16\pi s}\frac{{\rm p}^\prime}{\rm p}\sum_{J^P,\lambda'\geq0\lambda\geq0}\frac{\tilde{J}}{\tilde{j_1}\tilde{j_2}}\left|\frac{M^{J^P}_{\lambda'\lambda}({\rm p}', {\rm p})}{4\pi}\right|^2,
\end{eqnarray}
where the $s$ is the invariant mass square of the system of initial particles. 
$J$ and $j_i$ are total angular momentum of system and spins of  two initial
particles, respectively.

\section{Numerical results}\label{NUMERICAL RESULTS}

\subsection{Bound states from $\Lambda_c\bar{\Lambda}_c$ interaction}

With the above preparation, the possible molecular states from the
$\Lambda_c\bar{\Lambda}_c$ interaction can be searched in the complex energy
plane by adjusting the parameter $\alpha$. As usual, only the states from the
S-wave interaction are considered in the current work, that is, states with
quantum numbers $J^{PC}=0^{-+}$ and $1^{--}$. The single-channel calculation is
first performed, which produce bound state as a pole at real axis below the
threshold. The results are shown in Table~\ref{diagrams}.

\renewcommand\tabcolsep{0.42cm}
  \renewcommand{\arraystretch}{1.}
  \begin{table}[h!]\footnotesize
  \begin{center}
  \caption{The binding energies of bound states from the $\Lambda_c\bar{\Lambda}_c$ interaction at different cutoffs $\alpha$ in a range from 0.0 to 0.5.  Here, ``$--$" means that no bound state is found, or the bound state has a binding energy larger than 30 MeV.\label{Tab: bound state}
  \label{diagrams}}
  \begin{tabular}{ccccccc}\bottomrule[1pt]

  $\alpha$    &0.0        &$0.1$             & $0.2$            & $0.3$        & $0.4$    &$0.5$      \\\hline
  $0^{-+}$   &$3.0$         & $6.7$           &$11.8$            & $18.6$         &$26.8$       &$--$ \\
  $1^{--}$  &$3.0$             & $6.7$        &$11.8$            & $18.6$          &$26.8$      &$--$
  \\\toprule[1pt]
  \end{tabular}
  \end{center}
\end{table}

As listed in Table~\ref{diagrams}, the bound states can be produced from the
$\Lambda_c\bar{\Lambda}_c$ interaction  with both $J^{PC}=0^{-+}$ and $1^{--}$.
The two bound states have almost the  same binding energy, which is consistent
with the result in Ref.~\cite{Lee:2011rka}. The results indicate that the
attraction between $\Lambda_c$ baryon and $\bar{\Lambda}_c$ baryon is
considerably strong, which leads to first appearance of bound states at a value
of $\alpha$ below 0.  The binding energies of the bound states increase
gradually with increasing of $\alpha$ value, and exceed 30~MeV at an $\alpha$
value about 0.5.

\begin{figure}[h!]
  \centering
  \includegraphics[scale=0.85]{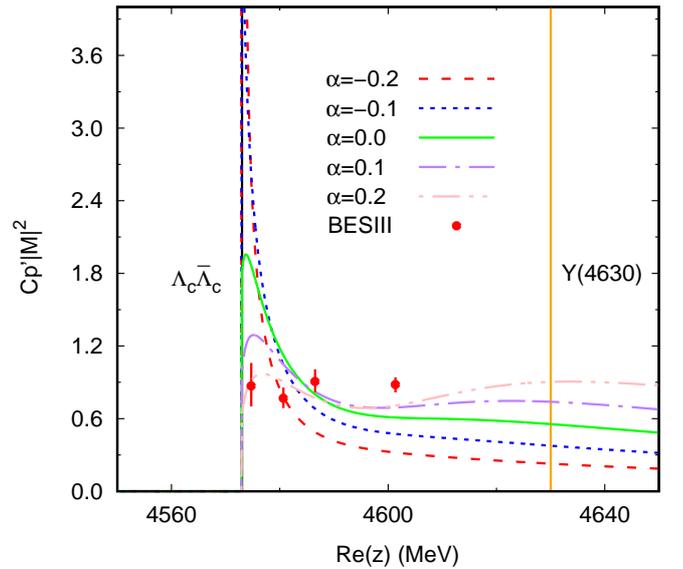}
  \caption{The $\Lambda_c\bar{\Lambda}_c$ invariant mass spectrum estimated by $C{\rm
  p}^\prime\left|M^{J^P}_{\lambda'\lambda}({\rm p}', {\rm p})\right|^2$ with $J^{PC}=1^{--}$ at $\alpha$= -0.2, -0.1, 0.0, 0.1 and 0.2. The orange line indicts the mass of Y(4630). The BESIII data for $e^+e^-\to \Lambda_c\bar{\Lambda}_c$ are cited from Ref.~\cite{BESIII:2017kqg}. \label{single01}}
\end{figure}

The bound state can be produced from the  $\Lambda_c\bar{\Lambda}_c$ interaction
with $J^{PC}=1^{--}$, which are also the quantum number of the Y(4630). As shown
in  Table~\ref{diagrams}, its binding energy is below the threshold even with an
$\alpha$ value of 0. However, the experimental structure of $Y(4630)$ is  about
60~MeV above the $\Lambda_c\bar{\Lambda}_c$ threshold. To discuss the relation
between the vector $\Lambda_c\bar{\Lambda}_c$ molecular state to the
experimentally observed $Y(4630)$, we make an estimation of the
$\Lambda_c\bar{\Lambda}_c$ invariant mass spectrum as $C{\rm
p}^\prime\left|M^{J^P}_{\lambda'\lambda}({\rm p}', {\rm p})\right|^2$ shown in
Fig.~\ref{single01} with different $\alpha$ values. Here $C$ is the Gamov-Sommerfeld factor for the Coulomb enhancement effect as $C=y/(1-e^{-y})$ with $y=\pi\alpha\sqrt{1-\beta^2}/\beta$ and $\beta=\sqrt{1-4m^2_{\Lambda_c}/W^2}$~\cite{Huang:2021xte,Sommerfeld:1931qaf}. 

With  $\alpha$ values of 0, a peak can be observed near the
$\Lambda_c\bar{\Lambda}_c$ threshold, which corresponds to a bound state with a
binding energy about 3~MeV. However, it is found quite difficult to produce a
peak near the mass about 4630 MeV even after varying the $\alpha$ value in
reasonable range. With the increase of  $\alpha$ value, the pole of the bound
state will leave the threshold further, which results in a smaller peak near the
threshold. If smaller $\alpha$ value is chosen, the bound state will disappear,
and no virtual and resonance can be produced in our model.  The peak is still on
the threshold. Hence, though a bound state can be produced from the
$\Lambda_c\bar{\Lambda}_c$ interaction with the same quantum numbers as
$Y(4630)$, it is difficult to be used to interpret the experimental observed
structure. BESIII reported the cross section of process $e^+e^-\to
\Lambda_c\bar{\Lambda}_c$ near the $\Lambda_c\bar{\Lambda}_c$
threshold~\cite{BESIII:2017kqg}, which are also  presented in
Fig.~\ref{single01} for reference. Here, the invariant mass spectrum is
estimated by the single-channel amplitude of process
$\Lambda_c\bar{\Lambda}_c\to\Lambda_c\bar{\Lambda}_c$, and can not be compared
directly. However, the rapid  increase can be found in both theoretical and
experimental results, which suggests that the $\Lambda_c\bar{\Lambda}_c$
interaction plays important role near the threshold.

\subsection{Production of $1^{--}$ state in nucleon-antinucleon collision}

\begin{figure}[h!]
\centering
\includegraphics[scale=0.7]{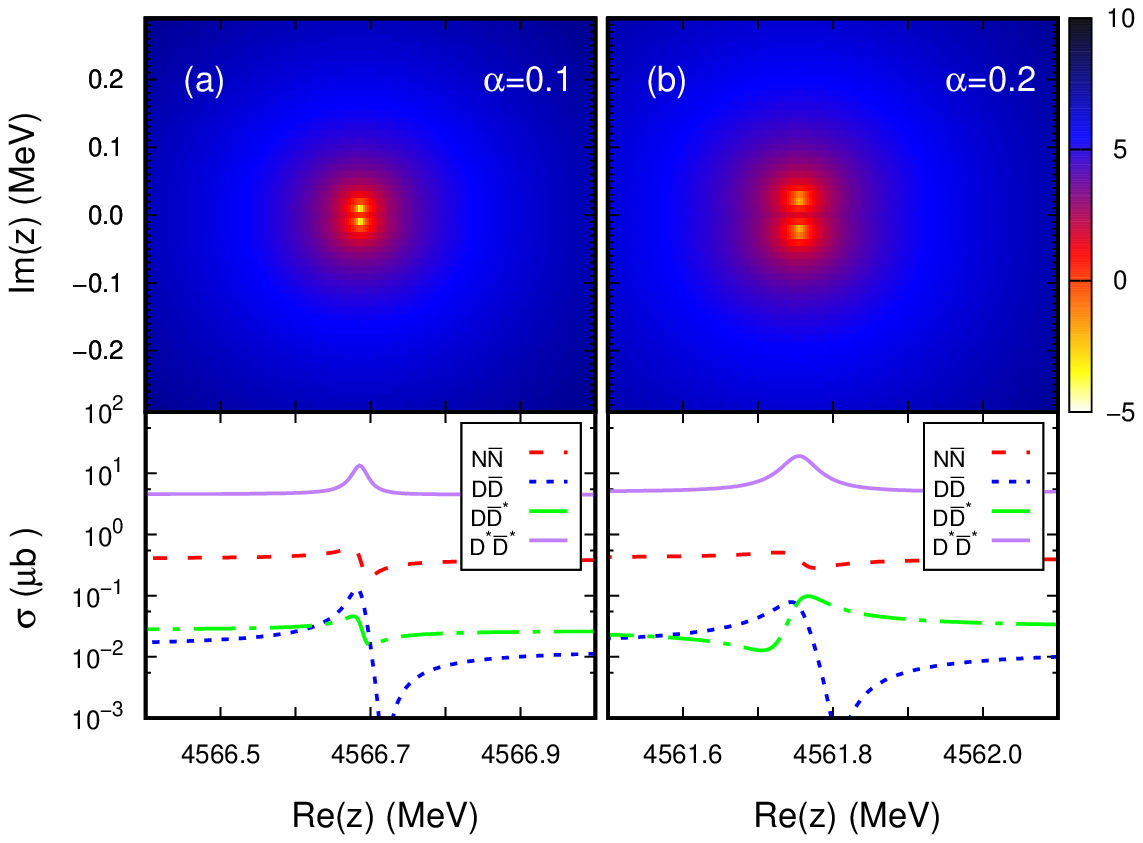}
\includegraphics[scale=0.7]{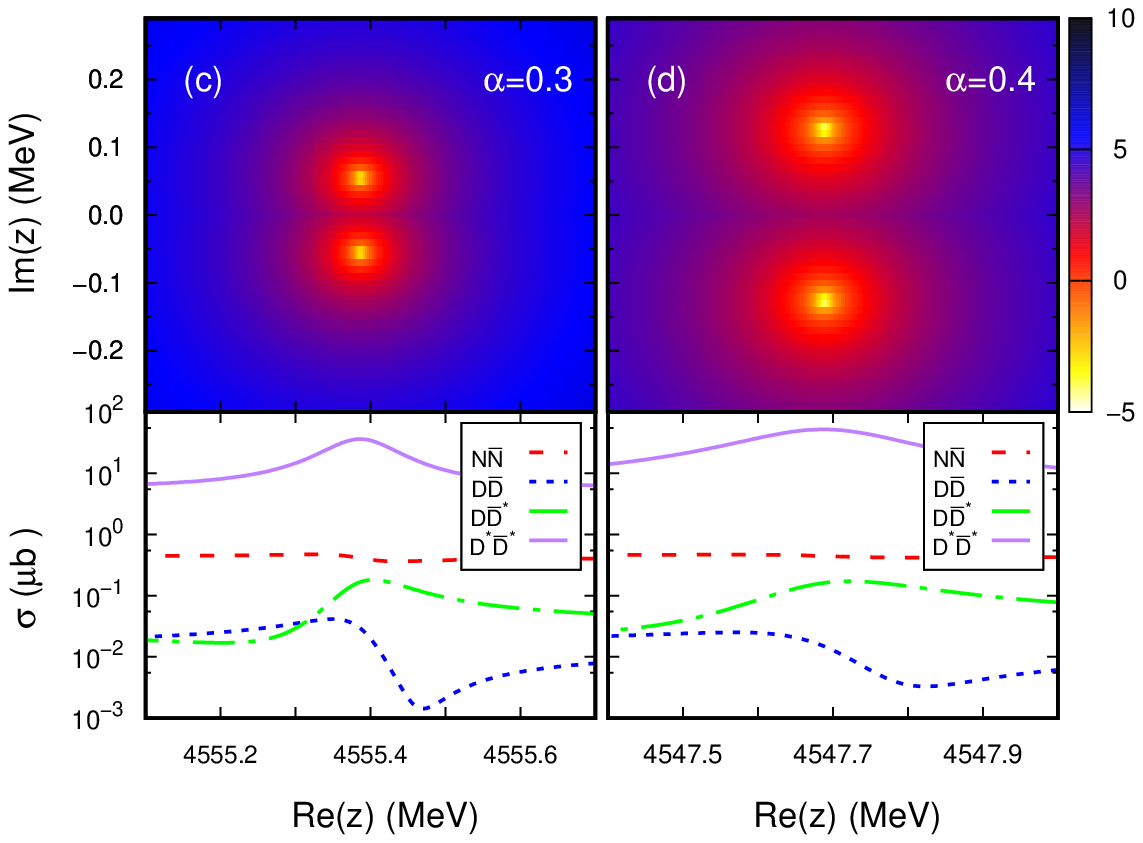}
\caption{The pole of bound state (upper) and the cross section (lower) from the $N\bar{N}$ collision with $J^{PC}=1^{--}$ at $\alpha$ = 0.1, 0.2, 0.3 and 0.4. The color  means the value of $\log|1-V(z)G(z)|$ as shown in the color box.\label{coupled01}}
\end{figure}

Now that vector molecular states produced from the $\Lambda_c\bar{\Lambda}_c$
interaction is difficult to be explained as the $Y(4630)$. In the following, we
discuss the possibility to observe it in the nucleon-antinucleon collision.
After the $N\bar{N}$ and $D^{(*)}\bar{D}^{(*)}$ channels and their coupling are
included, the cross section of production of $\Lambda_c\bar{\Lambda}_c$
molecular state  with $J^{PC}=1^{--}$ in the  nucleon-antinucleon collision is calculated, and the
results are illustrated in Fig~\ref{coupled01}.

Due to lack of experimental data, the cross section will be discussed with
different $\alpha$ values. Considering the uncertainties introduced by this
parameter, we will provide the results for the $N\bar{N}$ collision instead of
the explicit $p\bar{p}$ or $n\bar{n}$ collisions, which should have cross
section at the same order of magnitude. Here, we still focus on the energy
region near the $\Lambda_c\bar{\Lambda}_c$ threshold.  With the coupled-channel
effects are included, the pole for the molecular state leaves the real axis
of the complex energy plane. With the increase of $\alpha$ value, the
deviation from the real axis becomes larger, which means that the molecular
state has a larger width. Generally speaking, the width of the molecular state
with $J^{PC}=1^{--}$ is considerably small, at an order of magnitude of 0.1~MeV.

In the calculation, four channels, $N\bar{N}$, $D\bar{D}$, $D\bar{D}^*$, and
$D^*\bar{D}^*$, are considered as the final states. There are large difference
between cross sections of different channels.  The largest cross section can be
found in the  $D^{*}\bar{D}^*$ channel at an order of magnitude of 10~$\mu$b at
$\alpha$ values from 0.1 to 0.4.  Here, to show the results more explicitly,
energy ranges are chosen very small in Fig~\ref{coupled01}. We would like to
emphasize that the width of the state is very small, the peak should be very
sharp.  The cross section in the $N\bar{N}$ channel is much smaller than the
$D^{*}\bar{D}^*$ channel, at an order of magnitude of 0.1~$\mu$b. No obvious
peak can be found in this channel, but with a small structure due to the
interference between the contribution of molecular state and background.
Similar structures can be found in the lineshapes of cross sections in other two
channels, $D\bar{D}$ and $D\bar{D}^*$, at an order of magnitude of 0.01~$\mu$b.
Hence, both the height of the peak and the lineshapes of the cross sections
suggest that the $D^*\bar{D}^*$ channel is the best channel to search the
$\Lambda_c\bar{\Lambda}_c$  molecular state with $J^{PC}=1^{--}$.

\subsection{Production of $0^{-+}$ state in nucleon-antinucleon collision}

\begin{figure}[h!]
\centering
\includegraphics[scale=0.7]{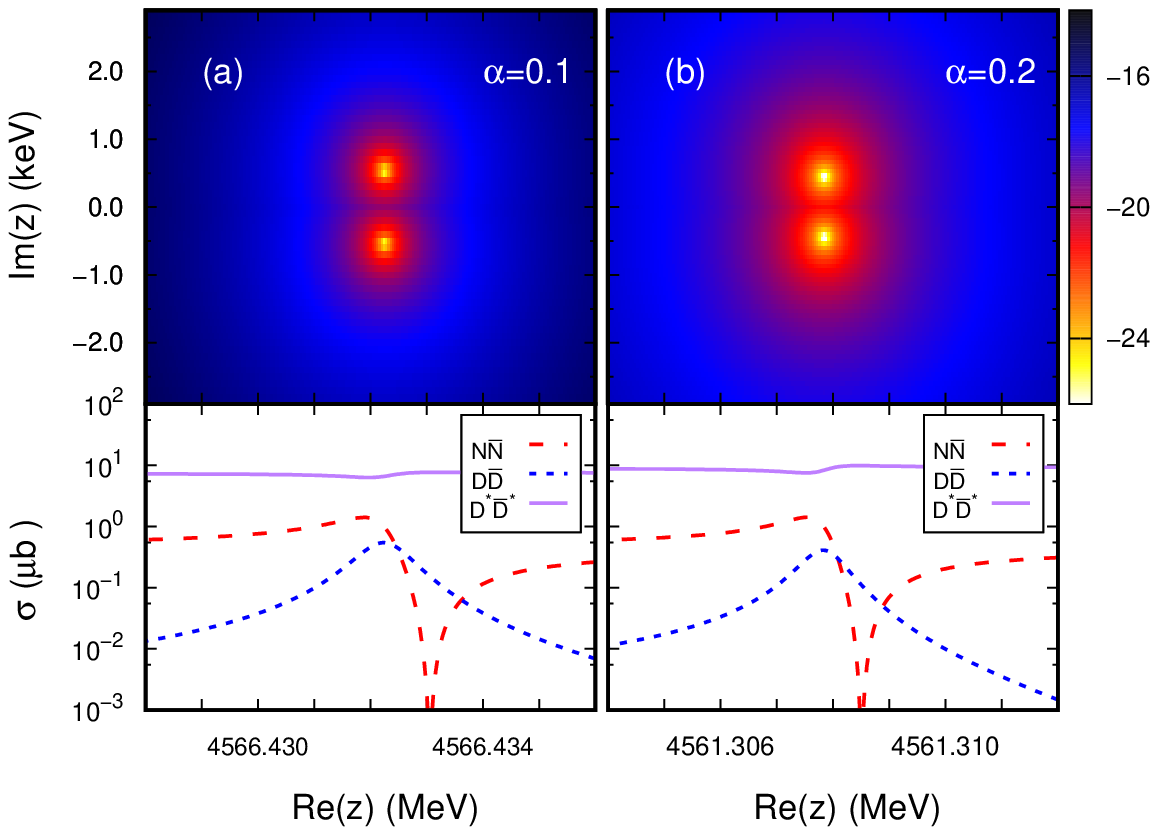}
\includegraphics[scale=0.7]{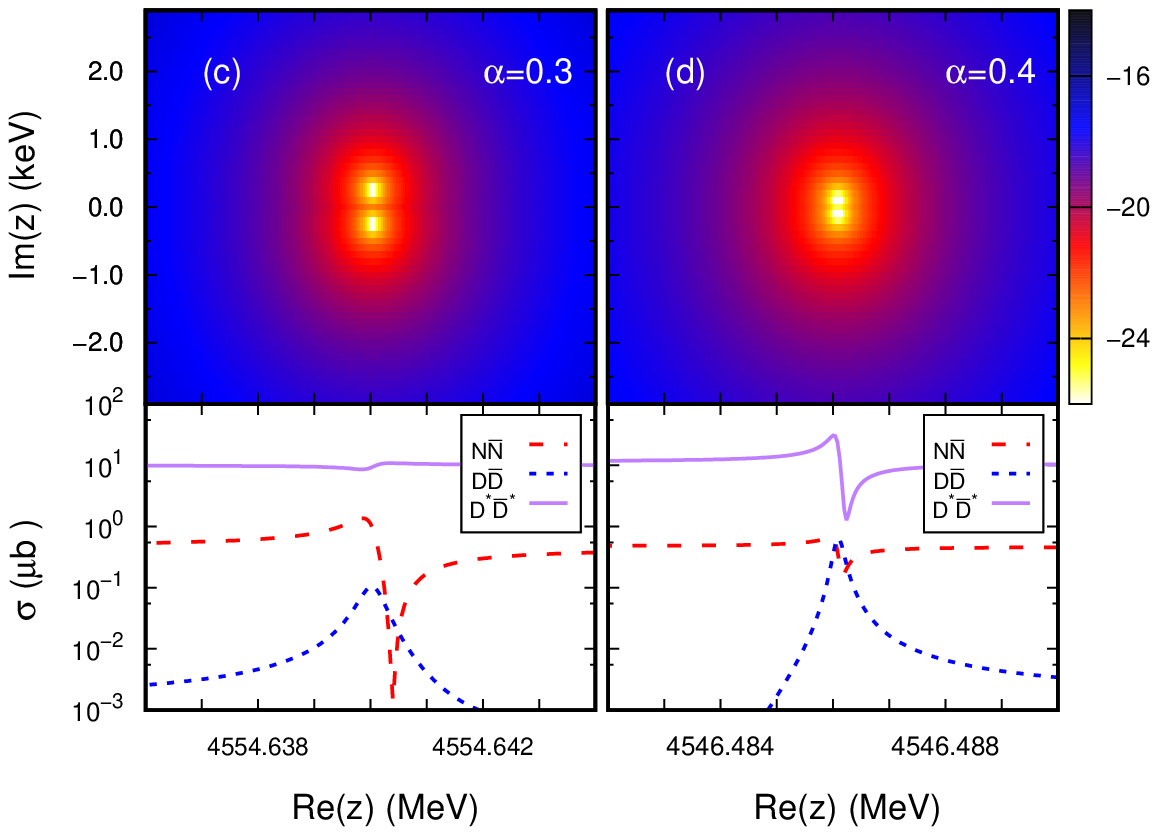}
\caption{The pole of bound state (upper) and the cross section (lower) from the interaction of $N\bar{N}$ collision with $J^{PC}=0^{-+}$ at $\alpha$ = 0.1, 0.2, 0.3 and 0.4. The color  means the value of $\log|1-V(z)G(z)|$ as shown in the color box.\label{coupled00}}
\end{figure}

The molecular state with $0^{-+}$ is also produced from the
$\Lambda_c\bar{\Lambda}_c$ interaction in S wave.  The pole of this state with
coupled-channel calculation and its production from the nucleon-antinucleon
collision are shown in Fig~\ref{coupled00}.  Since the decay of  state with
$0^{-+}$ to $D\bar{D}$ is forbidden, three channels, $N\bar{N}$,  $D\bar{D}^*$,
and $D^*\bar{D}^*$, are considered as final states.

Although two bound states produced in S wave are almost degenerated, the
molecular state with $0^{-+}$ exhibits different behaviors after the
coupled-channel effects are included. A very small width about $0.1$ KeV is
found in our calculation, which is 3 orders of magnitude lower than state with
$1^{--}$.  It reflects that the couplings of the state with $0^{-+}$ to the
final channels considered are very weak. The cross section in the $D^*\bar{D}^*$
channel is still at an order of magnitude of 10~$\mu$b, but the structure can
not be seen obviously at small $\alpha$ values. Only a structure from
interference can be found at $\alpha$ value of 0.4. A dip and peak can be found
in the channels $N\bar{N}$ and $D\bar{D}$, respectively, but with small
magnitudes. The structures in these two channels are obvious, but should
be very narrow due to the extremely small width of the molecular state.  It will
result in that the number of the events from these structures is very
small. Hence, the existence of the possible state with $0^{-+}$ may be not easy
to confirmed in experiment.

\section{Summary}\label{Summary}

In this work, the possible molecular states from the $\Lambda_c\bar{\Lambda}_c$
interaction are studied in a qBSE approach. The potential kernel is constructed
by the $\omega$ and $\sigma$ exchanges, which provide strong  attraction between
two baryons, and two bound states with quantum numbers $J^{PC}=1^{--}$ and
$0^{-+}$ are produced from the interaction.  Though it is alluring to interpret
the experimentally observed $Y(4630)$ as a $\Lambda_c\bar{\Lambda}_c$ molecular
state, the mass gap between the $Y(4630)$ and the molecular state, which is more
than 60 MeV, makes it hardly to fit the peak of molecular state to the
experimental structure above the threshold. Hence, it is difficult to explain
the $Y(4630)$ as a $\Lambda_c\bar{\Lambda}_c$ molecular state.

A coupled-channel calculation with $N\bar{N}$ and $D^{(*)}\bar{D}^{(*)}$
channels is performed to study the possibility to observe the molecular states
in the nucleon-antinucleon collision. The coupled-channel effect give the width
to two almost degenerated bound states in the single-channel calculation, which
makes two states behave in different manners. The structure appears near the
$\Lambda_c\bar{\Lambda}_c$  threshold due to the existence of the molecular
state.  For the molecular state with $J^{PC}=1^{--}$, a peak can be found
obviously in the $D^*\bar{D}^*$ channel at an order of magnitude of 10~$\mu$b.
Combined with a width of about 0.1~MeV, it is hopeful to search for the
molecular states in the $D^*\bar{D}^*$ channel. For state with $0^{-+}$, the
large cross section is also found in the $D^*\bar{D}^*$ channel, but no obvious
peak structure appears.  Dip and peak structures can be also found in the
channel $N{N}$ and $D\bar{D}$ with small magnitudes.  However, the width of
state with $0^{-+}$ is too small, which may lead to the difficulty in
experiment. Even if the contributions from the channels, which are not
considered in the current work, provide a larger width for this state, its
couplings to the channels considered in the current work may be too small to
provide enough events.  We provide the results for  $1^{--}$ and $0^{-+}$
states, respectively. The cross section can be obtained by adding the
contributions from two states directly.   Since the $N\bar{N}$ can couple to
both quantum numbers and the $\Lambda_c \bar{\Lambda}_c$ pair are in S wave for
both cases, the $N\bar{N}$ collisions will produce the states with both quantum
numbers simultaneously. Without the partial wave analysis, it will make the
structures in this energy region  more complex. However, for the process
$N\bar{N}\to D^*\bar{D}^*$, the $1^{--}$ state can stand out the background
because the contribution from the $0^{-+}$ state is relatively flat.  Hence,
based on the results in the current work, we suggest search for the
$\Lambda_c\bar{\Lambda}_c$ molecular state with $1^{--}$ at process $N\bar{N}\to
D^*\bar{D}^*$, which is accessible at $\rm \bar{P}ANDA$.

\section*{Acknowledgments}

This project is  supported by the National
Natural Science Foundation of China (Grant No.11675228).


\begin{thebibliography}{99}


\bibitem{Belle:2008xmh}
G.~Pakhlova \textit{et al.} [Belle],
``Observation of a near-threshold enhancement in the $e^+e^- \to \Lambda^+_c \Lambda^-_c$ cross section using initial-state radiation,''
Phys. Rev. Lett. \textbf{101}, 172001 (2008)

\bibitem{BESIII:2017kqg}
M.~Ablikim \textit{et al.} [BESIII],
``Precision measurement of the $e^{+}e^{-}~\rightarrow~\Lambda_{c}^{+} \bar{\Lambda}_{c}^{-}$ cross section near threshold,''
Phys. Rev. Lett. \textbf{120}, no.13, 132001 (2018)

\bibitem{Badalian:2008dv} 
A.~M.~Badalian, B.~L.~G.~Bakker and I.~V.~Danilkin,
``The S - D mixing and di-electron widths of higher charmonium $1^{--}$ states,''
Phys. Atom. Nucl. \textbf{72}, 638-646 (2009)

\bibitem{Segovia:2008ta}   
J.~Segovia, D.~R.~Entem and F.~Fernandez,
``Charm spectroscopy beyond the constituent quark model,''
[arXiv:0810.2875 [hep-ph]].

\bibitem{Xiao:2018iez}   
L.~Y.~Xiao, X.~Z.~Weng, Q.~F.~L\"u, X.~H.~Zhong and S.~L.~Zhu,
``A new decay mode of higher charmonium,''
Eur. Phys. J. C \textbf{78}, no.7, 605 (2018)


\bibitem{Wang:2020prx}  
J.~Z.~Wang, R.~Q.~Qian, X.~Liu and T.~Matsuki,
``Are the $Y$ states around 4.6 GeV from $e^+e^-$ annihilation higher charmonia?,''
Phys. Rev. D \textbf{101}, no.3, 034001 (2020)

\bibitem{Liu:2016sip}  
X.~Liu, H.~W.~Ke, X.~Liu and X.~Q.~Li,
``Exploring open-charm decay mode $\Lambda _c\bar{\Lambda }_c$ of charmonium-like state $Y(4630)$,''
Eur. Phys. J. C \textbf{76}, no.10, 549 (2016)

\bibitem{Wang:2018rfw}   
Z.~G.~Wang,
``Vector tetraquark state candidates: $Y(4260/4220)$, $Y(4360/4320)$, $Y(4390)$ and $Y(4660/4630)$,''
Eur. Phys. J. C \textbf{78}, no.6, 518 (2018)

\bibitem{vanBeveren:2008rt}  
E.~van Beveren, X.~Liu, R.~Coimbra and G.~Rupp,
``Possible $\psi(5S), \psi(4D), \psi(6S) and \psi(5D)$ signals in $\Lambda_c$ $\bar{\Lambda}_c$,''
EPL \textbf{85}, no.6, 61002 (2009)

\bibitem{Chen:2011cta}   
Y.~D.~Chen and C.~F.~Qiao,
``Baryonium Study in Heavy Baryon Chiral Perturbation Theory,''
Phys. Rev. D \textbf{85}, 034034 (2012)

\bibitem{Wang:2021qmn}
X.~W.~Wang, Z.~G.~Wang and G.~l.~Yu,
``Study of $\Lambda _c\Lambda _c$ dibaryon and $\Lambda _c{\bar{\Lambda }}_c$ baryonium states via QCD sum rules,''
Eur. Phys. J. A \textbf{57}, no.9, 275 (2021)


\bibitem{Lee:2011rka}  
N.~Lee, Z.~G.~Luo, X.~L.~Chen and S.~L.~Zhu,
``Possible Deuteron-like Molecular States Composed of Heavy Baryons,''
Phys. Rev. D \textbf{84}, 014031 (2011)

\bibitem{Simonov:2011jc}  
Y.~A.~Simonov,
``Theory of hadron decay into baryon-antibaryon final state,''
Phys. Rev. D \textbf{85}, 105025 (2012)

\bibitem{Belle:2007umv}  
X.~L.~Wang \textit{et al.} [Belle],
``Observation of Two Resonant Structures in $e^+e^- \to \pi^+ \pi^- \psi(2S)$ via Initial State Radiation at Belle,''
Phys. Rev. Lett. \textbf{99}, 142002 (2007)

\bibitem{Bugg:2008sk}  
D.~V.~Bugg,
``An Alternative fit to Belle mass spectra for $D\bar{D}, D^*\bar{D}^*$ and $\Lambda_C \bar{\Lambda}_c$,''
J. Phys. G \textbf{36}, 075002 (2009)

\bibitem{Cotugno:2009ys} 
G.~Cotugno, R.~Faccini, A.~D.~Polosa and C.~Sabelli,
``Charmed Baryonium,''
Phys. Rev. Lett. \textbf{104}, 132005 (2010)

\bibitem{Guo:2008zg}
F.~K.~Guo, C.~Hanhart and U.~G.~Meissner,
Phys. Lett. B \textbf{665}, 26-29 (2008)
doi:10.1016/j.physletb.2008.05.057
[arXiv:0803.1392 [hep-ph]].

\bibitem{Guo:2010tk}  
F.~K.~Guo, J.~Haidenbauer, C.~Hanhart and U.~G.~Meissner,
``Reconciling the X(4630) with the Y(4660),''
Phys. Rev. D \textbf{82}, 094008 (2010)


\bibitem{Dong:2021juy}  
X.~K.~Dong, F.~K.~Guo and B.~S.~Zou,
``A survey of heavy-antiheavy hadronic molecules,''
Progr. Phys. \textbf{41}, 65-93 (2021)

\bibitem{BESIII:2017kqg}
M.~Ablikim \textit{et al.} [BESIII],
``Precision measurement of the $e^{+}e^{-}~\rightarrow~\Lambda_{c}^{+} \bar{\Lambda}_{c}^{-}$ cross section near threshold,''
Phys. Rev. Lett. \textbf{120}, no.13, 132001 (2018)
Copy to ClipboardDownload


\bibitem{Titov:2008yf}  
A.~I.~Titov and B.~Kampfer,
``Exclusive charm production in $\bar{p}p$ collisions at $\sqrt{s}$ $\lesssim$ 15GeV,''
Phys. Rev. C \textbf{78}, 025201 (2008)

\bibitem{Goritschnig:2009sq} 
A.~T.~Goritschnig, P.~Kroll and W.~Schweiger,
``Proton-Antiproton Annihilation into a $\Lambda^+_c \bar{\Lambda}^-_c$ Pair,''
Eur. Phys. J. A \textbf{42}, 43-62 (2009)

\bibitem{Haidenbauer:2009ad}  
J.~Haidenbauer and G.~Krein,
``The Reaction $\bar{p}p\to\bar{\Lambda}^-_c\Lambda^+_c$ close to threshold,''
Phys. Lett. B \textbf{687}, 314-319 (2010)

\bibitem{Wang:2016fhj}    
Y.~Y.~Wang, Q.~F.~L\"u, E.~Wang and D.~M.~li,
``Role of $Y(4630)$ in the $p\bar{p}\rightarrow\Lambda_c\bar{\Lambda}_c$ reaction near threshold,''
Phys. Rev. D \textbf{94}, 014025 (2016)

\bibitem{Guo:2016iej}  
X.~D.~Guo, D.~Y.~Chen, H.~W.~Ke, X.~Liu and X.~Q.~Li,
``Study on the rare decays of $Y(4630)$ induced by final state interactions,''
Phys. Rev. D \textbf{93}, no.5, 054009 (2016)

\bibitem{Wiedner:2011mf}  
U.~Wiedner,
``Future Prospects for Hadron Physics at $\bar{P}ANDA$,''
Prog. Part. Nucl. Phys. \textbf{66}, 477-518 (2011)

\bibitem{Khodjamirian:2011sp} 
A.~Khodjamirian, C.~Klein, T.~Mannel and Y.~M.~Wang,
``How much charm can $\bar{P}ANDA$ produce?,''
Eur. Phys. J. A \textbf{48}, 31 (2012)

\bibitem{Shyam:2016bzq}
R.~Shyam and K.~Tsushima,
``${\bar D}D$ meson pair production in antiproton-nucleus collisions,''
Phys. Rev. D \textbf{94}, no.7, 074041 (2016)

\bibitem{Shi:2021hzm}
P.~P.~Shi, Z.~H.~Zhang, F.~K.~Guo and Z.~Yang,
``D+D- hadronic atom and its production in pp and pp\textasciimacron{} collisions,''
Phys. Rev. D \textbf{105}, no.3, 034024 (2022)


\bibitem{He:2019ify}  
J.~He,
``Study of $P_c(4457)$, $P_c(4440)$, and $P_c(4312)$ in a quasipotential Bethe-Salpeter equation approach,''
Eur. Phys. J. C \textbf{79}, no.5, 393 (2019)

\bibitem{Zhu:2019ibc} 
J.~T.~Zhu, Y.~Liu, D.~Y.~Chen, L.~Jiang and J.~He,
``$X$(2239) and $\eta (2225)$ as hidden-strange molecular states from $\Lambda \bar \Lambda$ interaction,''
Chin. Phys. C \textbf{44}, no.12, 123103 (2020)

\bibitem{He:2019csk}  
J.~He, Y.~Liu, J.~T.~Zhu and D.~Y.~Chen,
``Y(4626) as a molecular state from interaction ${D}^*_s{\bar{D}}_{s1}(2536)-{D}_s{\bar{D}}_{s1}(2536)$,''
Eur. Phys. J. C \textbf{80}, no.3, 246 (2020)

\bibitem{Yan:1992gz}  
T.~M.~Yan, H.~Y.~Cheng, C.~Y.~Cheung, G.~L.~Lin, Y.~C.~Lin and H.~L.~Yu,
``Heavy quark symmetry and chiral dynamics,''
Phys. Rev. D \textbf{46}, 1148-1164 (1992)

\bibitem{Cheng:1992xi}  
H.~Y.~Cheng, C.~Y.~Cheung, G.~L.~Lin, Y.~C.~Lin, T.~M.~Yan and H.~L.~Yu,
``Chiral Lagrangians for radiative decays of heavy hadrons,''
Phys. Rev. D \textbf{47}, 1030-1042 (1993)

\bibitem{Casalbuoni:1996pg} 
R.~Casalbuoni, A.~Deandrea, N.~Di Bartolomeo, R.~Gatto, F.~Feruglio and G.~Nardulli,
``Phenomenology of heavy meson chiral Lagrangians,''
Phys. Rept. \textbf{281}, 145-238 (1997)


\bibitem{He:2019rva}  
J.~He and D.~Y.~Chen,
``Molecular states from $\Sigma^{(*)}_c\bar{D}^{(*)}-\Lambda_c\bar{D}^{(*)}$ interaction,''
Eur. Phys. J. C \textbf{79}, no.11, 887 (2019)

\bibitem{Zhu:2021lhd}  
J.~T.~Zhu, L.~Q.~Song and J.~He,
``$P_{cs}(4459)$ and other possible molecular states from $\Xi_{c}^{(*)}\bar{D}^{(*)}$ and $\Xi'_c\bar{D}^{(*)}$ interactions,''
Phys. Rev. D \textbf{103}, no.7, 074007 (2021)

\bibitem{Ronchen:2012eg}  
D.~Ronchen, M.~Doring, F.~Huang, H.~Haberzettl, J.~Haidenbauer, C.~Hanhart, S.~Krewald, U.~G.~Meissner and K.~Nakayama,
``Coupled-channel dynamics in the reactions $\pi N\to\pi N, \eta N, K\Lambda, K\Sigma$,''
Eur. Phys. J. A \textbf{49}, 44 (2013)

\bibitem{Matsuyama:2006rp}
A.~Matsuyama, T.~Sato and T.~S.~H.~Lee,
``Dynamical coupled-channel model of meson production reactions in the nucleon resonance region,''
Phys. Rept. \textbf{439}, 193-253 (2007)

\bibitem{Lu:2020qme}    
Z.~T.~Lu, H.~Y.~Jiang and J.~He,
``Possible molecular states from the $N\Delta$ interaction,''
Phys. Rev. C \textbf{102}, no.4, 045202 (2020)

\bibitem{Phillips:1967zza} 
R.~j.~n.~Phillips,
``Antinuclear Forces,''
Rev. Mod. Phys. \textbf{39}, 681-688 (1967)

\bibitem{Aceti:2014uea}  
F.~Aceti, M.~Bayar, E.~Oset, A.~Martinez Torres, K.~P.~Khemchandani, J.~M.~Dias, F.~S.~Navarra and M.~Nielsen,
``Prediction of an $I=1$ $D \bar D^*$ state and relationship to the claimed $Z_c(3900)$, $Z_c(3885)$,''
Phys. Rev. D \textbf{90}, no.1, 016003 (2014)

\bibitem{He:2017lhy}  
J.~He and D.~Y.~Chen,
``$Z_c(3900)/Z_c(3885)$ as a virtual state from $\pi J/\psi-\bar{D}^*D$ interaction,''
Eur. Phys. J. C \textbf{78}, no.2, 94 (2018)

\bibitem{Okubo:1975sc}  
S.~Okubo,
``SU(4), SU(8) Mass Formulas and Weak Interactions,''
Phys. Rev. D \textbf{11}, 3261-3269 (1975)

\bibitem{Sibirtsev:2000aw} 
A.~Sibirtsev, K.~Tsushima and A.~W.~Thomas,
``Charmonium absorption by nucleons,''
Phys. Rev. C \textbf{63}, 044906 (2001)

\bibitem{Liu:2001ce}  
W.~Liu, C.~M.~Ko and Z.~W.~Lin,
``Cross-section for charmonium absorption by nucleons,''
Phys. Rev. C \textbf{65}, 015203 (2002)

\bibitem{Dong:2009tg}  
Y.~Dong, A.~Faessler, T.~Gutsche and V.~E.~Lyubovitskij,
``Strong two-body decays of the $\Lambda_c(2940)^+$ in a hadronic molecule picture,''
Phys. Rev. D \textbf{81}, 014006 (2010)

\bibitem{Ding:2020dio}  
Z.~M.~Ding, H.~Y.~Jiang and J.~He,
``Molecular states from $D^{(*)}\bar{D}^{(*)}/B^{(*)}\bar{B}^{(*)}$ and $D^{(*)}D^{(*)}/\bar{B}^{(*)}\bar{B}^{(*)}$ interactions,''
Eur. Phys. J. C \textbf{80}, no.12, 1179 (2020)

\bibitem{He:2015mja}  
J.~He,
``The $Z_c(3900)$ as a resonance from the $D\bar{D}^*$ interaction,''
Phys. Rev. D \textbf{92}, no.3, 034004 (2015)

\bibitem{He:2015cca}  
J.~He and P.~L.~Lu,
``The octet meson and octet baryon interaction with strangeness and the $\Lambda(1405)$,''
Int. J. Mod. Phys. E \textbf{24}, no.11, 1550088 (2015)

\bibitem{Huang:2021xte}
G.~Huang \textit{et al.} [BESIII],
``Probing the internal structure of baryons,''
Natl. Sci. Rev. \textbf{8}, no.11, nwab187 (2021)



\bibitem{Sommerfeld:1931qaf}
A.~Sommerfeld,
``\"Uber die Beugung und Bremsung der Elektronen,''
Annalen Phys. \textbf{403}, no.3, 257-330 (1931)


\bibitem{BESIII:2017kqg}
M.~Ablikim \textit{et al.} [BESIII],
``Precision measurement of the $e^{+}e^{-}~\rightarrow~\Lambda_{c}^{+} \bar{\Lambda}_{c}^{-}$ cross section near threshold,''
Phys. Rev. Lett. \textbf{120}, no.13, 132001 (2018)











\end{thebibliography}
\end{document}